\documentstyle[12pt]{article}
\def\frakd#1#2{{\displaystyle#1\over\displaystyle#2}}

\begin{document}
\flushbottom
\centerline{\Large \bf
On the Proton Form Factors in the Time-like Region\footnote[2]
{
Talk presented at the Workshop on Exclusive Reactions
at High Momentum Transfer,
24-26 June 1993, Marciana Marina, Elba, Italy.
}
}
\bigskip
\centerline{\large \sf
S.M. Bilenky$^{\mathrm{(a,b,c)}}$,
C. Giunti$^{\mathrm{(b,c)}}$
and
V. Wataghin$^{\mathrm{(b,c)}}$}
\medskip
\centerline{
(a) Joint Institute of Nuclear Research, Dubna, Russia }
\medskip
\centerline{
(b) INFN Torino, Via P. Giuria 1, I--10125 Torino, Italy }
\medskip
\centerline{
(c) Dipartimento di Fisica Teorica, Universit\`a di Torino }
\bigskip
\centerline{Abstract}
\begin{quote}
The process
$ \bar{p} p \rightarrow e^{-} e^{+} $
is considered
in the general case of polarized initial particles.
A relation between the difference of the phases
of the electromagnetic form factors $G_M$ and $G_E$
in the time-like region
and measurable asymmetries
is derived.
It is shown
that the moduli of the form factors
can be determined
from measurements
of the total unpolarized cross section
and of the integral asymmetry
for longitudinally polarized
(or transversely polarized)
$\bar{p}$ and $p$.
The behaviour of the proton form factors
at high $q^2$ in the time-like region
is also discussed.
{}From the
Phragm\'en-Lindel\"of's theorem
it follows
that the asymptotical behaviour of the form factors
in the space-like and time-like regions
must be the same.
An analysis of experimental data
in both regions
based on perturbative QCD is presented.
\end{quote}

\vspace{0.5cm}

The charge and magnetic form factors of the nucleon,
$ G_{E}(q^2) $
and
$ G_{M}(q^2) $,
are classical objects of investigation.
For a long time these fundamental quantities
have been investigated in the region of space-like $q^2$.
Starting from the seventies
some informations
on the electromagnetic form factors
of the proton in the time-like region
have been obtained.
Recently rather accurate measurements
of the proton form factors
in the time-like region,
from
$ q^2 = 4 M^2 $
up to
$ q^2 = 4.2 \,\mathrm{GeV}^2 $,
have been done at LEAR~\cite{B:LEAR91}.
Some informations
on the electromagnetic form factor of the proton
at high time-like $ q^2 $
were also obtained
at Fermilab~\cite{B:FNAL92}.

There exist several QCD calculations of the electromagnetic form factors
in the space-like region~\cite{B:QCD}.
According to our knowledge there are no QCD-based calculations
of the form factors in the time-like region.
The phenomenological models which try to describe
the behaviour of the form factors in both
space-like and time-like regions
are based on the vector meson dominance
models~\cite{B:VDM}.
However,
even taking into account all known meson resonances
it is not possible to obtain a statistically acceptable description
of all the existing experimental data.

The understanding of
the behaviour
of nucleon electromagnetic form factors
still remains a challenge for the theory.
It is clear that any additional information
about the form factors which could be obtained from experiment
is very important.

Taking into account
possible future developments
of the experiments at LEAR~\cite{B:LEAP92},
we have analyzed~\cite{B:BGW93}
which additional informations on
the proton form factors in the time-like region
can be obtained from the investigation of the process
\begin{equation}
\bar{p} p \rightarrow e^{-} e^{+}
\label{E1}
\end{equation}
with a polarized proton target
and/or a polarized antiproton beam.

The nucleon form factors in the time-like region
are complex.
In the case of unpolarized initial particles
the cross section depends only on the squared moduli
$|G_M|^2$ and $|G_E|^2$.
The study of process (\ref{E1})
with polarized initial particles
could allow to obtain informations
also about the phase difference
$ \chi = \chi_M - \chi_E $,
where
$ \chi_M = \mathrm{Arg} \, G_M $
and
$ \chi_E = \mathrm{Arg} \, G_E $,
which
is an important characteristic
of the form factors in the time-like region.

In ref.\cite{B:BGW93}
we gave the differential cross section of process (\ref{E1})
in the general case of polarized initial particles
and
we analyzed differential
and integral asymmetries.
Here we present some results
for the integral asymmetries.

The value of $\sin\chi$
can be obtained from measurements of the cross section
of process (\ref{E1})
with
an unpolarized antiproton beam
and
a polarized proton target
(or
a polarized antiproton beam
and
an unpolarized proton target).
If the target polarization
$ \vec{P}_{\perp} $
is orthogonal to the beam direction,
for the asymmetry
integrated over the angle $\varphi$
between
$ \vec{P}_{\perp} $
and
${\vec{\hskip+1pt{n}}\hskip+1pt}$
(the unit vector orthogonal to the reaction plane)
from
$ - \pi/2 $ to $ \pi/2 $
and over $\vartheta$
(the angle between the momenta of the antiproton and the electron
in the c.m.s.)
from $0$ to $\pi/2$
we have
\begin{equation}
\hbox{\sf A}_{\perp}
=
\frakd{
\frakd{ 4 M }{ \pi \sqrt{q^2} } \,
|G_M| |G_E| \sin\chi
}{
2 |G_M|^2
+
\frakd{ 4 M^2 }{ q^2 } \,
|G_E|^2
}
\;.
\label{E26}
\end{equation}
Information about $\cos\chi$
can be obtained from measurements
of the cross section of process (\ref{E1})
with
a transversely polarized beam
and a longitudinally polarized target
(or
a longitudinally polarized beam
and a transversely polarized target).
For the asymmetry
integrated
over the angle $\varphi'$ between $ \vec{P}'_{\perp} $
(the antiproton polarization vector)
and ${\vec{\hskip+1pt{n}}\hskip+1pt}$
from $0$ to $\pi$
and over the angle $\vartheta$
from $0$ to $\pi/2$
we have
\begin{equation}
\hbox{\sf A}_{\perp;\parallel}
=
\mbox{}
-
\frakd{
\frakd{ 4 M }{ \pi \sqrt{q^2} } \,
|G_M| |G_E| \cos\chi
}{
2 |G_M|^2
+
\frakd{ 4 M^2 }{ q^2 } \,
|G_E|^2
}
\;.
\label{E27}
\end{equation}

{}From Eqs.(\ref{E26}) and (\ref{E27})
we obtain the following relation between
the phase difference $\chi$
and the integral asymmetries
\begin{equation}
\tan\chi
=
\mbox{}
-
\frakd{ \hbox{\sf A}_{\perp} }{ \hbox{\sf A}_{\perp;\parallel} }
\;.
\label{E28}
\end{equation}
Thus measurements of the integral asymmetries
$ \hbox{\sf A}_{\perp} $
and
$ \hbox{\sf A}_{\perp;\parallel} $
would allow us to determine the phase difference $\chi$
directly from experimental data.
Notice that
the phase difference $\chi$
can be determined unambiguously
with this method
(in addition to Eq.(\ref{E28})
it is necessary to take into account the sign of
$ \hbox{\sf A}_{\perp} $
or
$ \hbox{\sf A}_{\perp;\parallel} $).
Let us stress that the knowledge
of the moduli
$|G_M|$ and $|G_E|$
is not necessary
in order to obtain $\chi$
with the help of Eq.(\ref{E28}).

Since
$ G_E(4M^2) = G_M(4M^2) $,
it is clear that the asymmetry
$ \hbox{\sf A}_{\perp} $
vanishes at the threshold.
It is possible to show that
the asymmetry
$ \hbox{\sf A}_{\perp} $
goes to zero at
$ q^2 \to \infty $.
In fact the electromagnetic form factors
$ G_{E,M}(q^2) $
are
limiting values of the functions
$ G_{E,M}(z) $,
$ \displaystyle
G_{E,M}(q^2)
=
\lim_{\epsilon\to0^{+}} G_{E,M}(q^2+i\epsilon)
$,
which are analytical in the upper half
of the complex $z$ plane
and increase at infinity not faster than a power of $z$.
We can apply~\cite{B:LHT65}
to the form factors
the Phragm\'en-Lindel\"of's theorem~\cite{B:PL}.
{}From this theorem it follows that
the form factors have the same asymptotical behaviour
in the space-like and time-like regions.
In the space-like region
the form factors are real.
This means that
the form factors are real in the time-like region
at asymptotically high $ q^2 $
and
from Eq.(\ref{E26})
it follows that
$ \hbox{\sf A}_{\perp} \to 0 $
at $ q^2 \to \infty $.

In conclusion,
let us discuss in some more detail
the asymptotical behaviour
of the electromagnetic form factors of the nucleon
in the time-like region.
In accordance with the quark counting rule
\cite{B:QCR},
at high $ |q^2| $
the form factors of the nucleon
behave as
$ G_{M}(q^2) \sim F_{1}(q^2) \sim 1/q^4 $.
The quark-gluon interaction
leads to violation of scaling
and additional logarithmic $q^2$ dependence of the form factors.
Let us write
in the space-like region ($q^2<0$)
\begin{equation}
\frakd{ G_{M}(q^2) }{ \mu_p }
\
\smash{\mathop{\sim}\limits_{q^2\to-\infty}}
\
\frakd{ C_s }{ q^4 } \, \Phi(q^2)
\;,
\label{E33}
\end{equation}
where $\mu_p=2.79$
is the proton magnetic moment
in nuclear magnetons.
{}From
the leading order perturbative QCD
it follows that
$ \Phi(q^2) $
is given by~\cite{B:BL}
\begin{equation}
\Phi(q^2)
=
\alpha_s^2(-q^2)
\left[ \ln\left(
- \frakd{ q^2 }{ \Lambda^2 }
\right) \right]^{-4/3\beta}
\;,
\qquad \mathrm{with} \qquad
\alpha_s(-q^2)
=
\frakd{ 4 \pi }{ \beta \ln\left( - \frakd{q^2}{\Lambda^2} \right) }
\;,
\label{E34}
\end{equation}
where
$ \beta = 11 - 2 n_f / 3 $,
$n_f$
is the number of flavours
and
$\Lambda$
is the QCD scale parameter.
The value of the constant
$C_s$
is determined by the wave function of the
nucleon~\cite{B:BL,B:CZ}.
The Phragm\'en-Lindel\"of's theorem implies
that
in the time-like region ($q^2>0$)
the form factors have the following asymptotical behaviour
\begin{equation}
\frakd{ G_{M}(q^2) }{ \mu_p }
\
\smash{\mathop{\sim}\limits_{q^2\to\infty}}
\
\frakd{ C_t }{ q^4 } \, \Phi(q^2)
\label{E36}
\end{equation}
with
$ C_t = C_s $.

Let us compare
the behaviour of the form factors
at large space-like and time-like
momentum transfer.
In a recent SLAC experiment~\cite{B:SLAC92}
the elastic electron-proton cross section
was measured in a wide range of momentum transfer,
from
$ -q^2 = 2.9 \,\mathrm{GeV}^2 $
to
$ -q^2 = 31.3 \,\mathrm{GeV}^2 $.
{}From these measurements
the values of the form factor $G_{M}(q^2)$
at high $-q^2$
can be extracted
(the contribution to the cross section
of the form factor $G_{E}(q^2)$
at high $-q^2$
is small and can be neglected).
With
$ \Lambda = 100 \,\mathrm{MeV} $
we obtain
$ C_s = 12.3 \pm 0.2 \,\mathrm{GeV}^4 $
($ \chi^2 / \mathrm{NDF} = 4.6 / 5 $)
and with
$ \Lambda = 200 \,\mathrm{MeV} $
we get
$ C_s = 7.8 \pm 0.1 \,\mathrm{GeV}^4 $
($ \chi^2 / \mathrm{NDF} = 10.2 / 5 $).
Let us notice that the quality of the fit
depends rather strongly
on the value of $\Lambda$
(smaller values of
$ \Lambda $
are preferable).

The cross section of the process
$ \bar{p} p \rightarrow e^{-} e^{+} $
at high $ q^2 $
($ q^2 = 8.9 $, $ 12.4 $, $ 13.0 \,\mathrm{GeV}^2 $)
was measured recently
in a Fermilab experiment~\cite{B:FNAL92}.
We made a fit of these data using Eq.(\ref{E36}).
For
$ \Lambda = 100 \,\mathrm{MeV} $
we obtained
$ C_t = 30.9^{+4.1}_{-4.8} \,\mathrm{GeV}^4 $
($ \chi^2 / \mathrm{NDF} = 0.29 / 1 $)
and for
$ \Lambda = 200 \,\mathrm{MeV} $
we obtained
$ C_t = 21.2^{+2.8}_{-3.3} \,\mathrm{GeV}^4 $
($ \chi^2 / \mathrm{NDF} = 0.29 / 1 $).
Thus the experimental data in the space-like
as well as in the time-like regions of $q^2$
are described by expressions (\ref{E33}) and (\ref{E36}),
respectively.
The accuracy of the data
in the time-like region
is much worse
than that
in the space-like region.
As a consequence,
the corresponding
accuracy of the determination
of the constant $C_t$
in the time-like region
is much worse
than that
of the constant $C_s$
in the space-like region.
However,
the average values of $C_t$ and $C_s$
are so different that,
even with such a low accuracy in the determination of $C_t$,
we can conclude
that these constants are different
($C_t$ is more than $3\sigma$ higher than $C_s$).
{}From our point of view
this difference means that
the range of high $q^2$ values
investigated in present experiments is not asymptotic.
Nonperturbative effects~\cite{B:ILR},
nonleading log corrections
and other effects
could be important
in this region.

\end{document}